\documentclass[12pt,preprint]{aastex}
\usepackage{graphicx}

\newcommand{\um}{$\mu$m}
\newcommand{\spitzer}{{\it Spitzer}}

\newcommand{\Rin}{$R_{\rm in}$}
\newcommand{\Rout}{$R_{\rm out}$}
\newcommand{\Rsub}{$R_{\rm sub}$}
\newcommand{\gcc}{g cm$^{-3}$}
\slugcomment{submitted to ApJ}
 
\shorttitle{FEPS: Cold Outer Disks}
\shortauthors{Kim et al.}
 
\begin{document}
 
\title{Formation and Evolution of Planetary Systems: \\ 
Cold Outer Disks Associated with Sun-like stars}

\author{
Jinyoung Serena Kim\altaffilmark{1},
Dean C. Hines\altaffilmark{2},
Dana E. Backman\altaffilmark{3}, 
Lynne A. Hillenbrand\altaffilmark{4},
Michael R. Meyer\altaffilmark{1},
Jens Rodmann\altaffilmark{5},
Amaya Moro-Mart{\'i}n\altaffilmark{6},
John M. Carpenter\altaffilmark{4},
Murray D. Silverstone\altaffilmark{1},
Jeroen Bouwman\altaffilmark{5},
Eric E. Mamajek\altaffilmark{7}, 
Sebastian Wolf\altaffilmark{4},
Renu Malhotra\altaffilmark{8},
Ilaria Pascucci\altaffilmark{1},
Joan Najita\altaffilmark{9},
Deborah L. Padgett\altaffilmark{10},
Thomas Henning\altaffilmark{5},
Timothy Y. Brooke\altaffilmark{4},
Martin Cohen\altaffilmark{11},
Stephen E. Strom\altaffilmark{9}, 
Elizabeth B. Stobie\altaffilmark{1}, 
Charles W. Engelbracht\altaffilmark{1}, Karl D. Gordon\altaffilmark{1},
Karl Misselt\altaffilmark{1}, Jane E. Morrison\altaffilmark{1},
James Muzerolle\altaffilmark{1}, \& Kate Y. L. Su\altaffilmark{1}
}

\altaffiltext{1}{Steward Observatory, The University of Arizona, 933 N. Cherry Ave., 
Tucson, AZ 85721-0065}
\altaffiltext{2}{Space Science Institute, 4750 Walnut Street, Suite 205, Boulder, CO}
\altaffiltext{3}{SOFIA, MS 211-3, NASA-Ames, Moffet Field, CA 94035-1000}
\altaffiltext{4}{Astronomy, California Institute of Technology, Pasadena, CA 91125}
\altaffiltext{5}{Max-Planck-Institut fur Astronomie, D-69117, Heidelberg, Germany}
\altaffiltext{6}{Princeton University, Princeton, NJ 08540}
\altaffiltext{7}{Harvard-Smithsonian Center for Astrophysics, 60 Garden St., 
MS-42 Cambridge, MA 02138}
\altaffiltext{8}{Department of Planetary Sciences \& Lunar and Planetary Laboratory, 
The University of Arizona, 1629 E. University Blvd., Tucson, AZ85721-0092}
\altaffiltext{9}{National Optical Astronomy Observatory, 950 N. Cherry Ave.,Tucson, AZ 85719} 
\altaffiltext{10}{Spitzer Science Center, California Institute of Technology, 
Pasadena, CA, 91125}
\altaffiltext{11}{Radio Astronomy, University of California, Berkeley, CA 94720}

\begin{abstract}

We present the discovery of debris systems around three solar mass stars 
based upon observations performed with the {\it Spitzer Space Telescope} 
as part of a Legacy Science Program, ``the Formation and Evolution        
of Planetary Systems'' (FEPS).  We also confirm the presence of debris      
around two other stars.  All the stars exhibit infrared emission in excess   
of the expected photospheres in the 70~$\mu$m band, but are consistent with   
photospheric emission at $\leq$33$~\mu$m. This restricts the maximum 
temperature of debris in equilibrium with the stellar radiation to 
$T < 70$~K. We find that these sources are relatively old in the FEPS 
sample, in the age range 0.7 $-$ 3 Gyr. Based on models of the spectral 
energy distributions, we suggest that these debris systems represent 
materials generated by collisions of planetesimal belts.
We speculate on the nature of these systems through comparisons to our own 
Kuiper Belt, and on the likely planet(s) responsible for 
stirring the system and ultimately releasing dust through collisions.  
We further report observations of a nearby star HD~13974 ($d =$ 11~pc) that 
is indistinguishable from a bare photosphere at both 24~\um\ and 70~\um.
The observations place strong upper limits on the presence of any 
cold dust in this nearby system ($L_{\rm IR}/L_\star < $10$^{-5.2}$).

\end{abstract}


\section{INTRODUCTION}

Studying the formation of our own solar system and observing the frequency 
of similar systems associated with other stars are two ways in which 
we seek to understand our origins.  Through remote observation and direct
exploration, we have developed a much clearer understanding of our
solar system. However many questions about the processes involved in the
initial formation and subsequent evolution towards the present
configuration cannot be addressed directly.  Therefore, we need to study
other stars to help place our solar system in context.

There are two major zones of debris in the solar system: the asteroid 
belt at 2 $-$ 4~AU composed of rocky material that is ground up by collisions 
to produce most of the zodiacal dust cloud and the Kuiper Belt (KB) that 
consists of small bodies orbiting beyond Neptune's orbit at 30 $-$ 50~AU.
Since their discovery over a decade ago (Jewitt \& Luu 1993), Kuiper Belt 
objects have played an increasingly important role in understanding 
the formation and evolution of our planetary system (e.g. Malhotra 1993; 
Kenyon \& Bromley 2004).   While direct detection of {\it in situ} debris 
from collisions amongst Kuiper Belt objects has yet to be confirmed, 
of order 10\% of the solar system's far-IR luminosity could be emitted by 
Kuiper Belt dust (e.g.\ Backman et al. 1995).
Searching for KB-like debris around other Sun-like stars as a function 
of age will determine both the frequency of such systems and
provide important insight into the formation of our solar system.

Most debris disks were found by their mid-to-far infrared emission in 
excess of the expected photosphere (e.g., Auman et al.\ 1984; Backman \& 
Paresce 1993; Mannings \& Barlow 1998; Habing et al.\ 1999; Silverstone 2000;
Spangler et al.\ 2001, Decin et al.\ 2003, Zuckerman \& Song 2004).  
The majority of these debris disks are also associated with hot,
luminous stars, since observatories such as IRAS and ISO did not have the
necessary sensitivities to detect debris disks around lower-luminosity
solar-type stars at distances beyond a few parsecs. 

The increased sensitivity afforded by the {\it Spitzer Space
Telescope} ({\it Spitzer}; Werner et al.\ 2004) has the potential to
identify and investigate debris systems that were not detectable with
previous observatories.  Using data from the Formation and Evolution
of Planetary Systems (FEPS) {\it Spitzer} Legacy Program, we have
conducted a search for outer dust disks dominated by temperatures 
characteristic of the Kuiper Belt ($T \sim 30 - 60$ K).  
The earliest results from our validation observations are presented in
Meyer et al.\ (2004), where we identified such debris disks (exo-KBs)
surrounding the 30 Myr-old Sun-like star HD 105 and the 0.4 - 1 Gyr
old star HD 150706.  
Herein we present five more Sun-like stars that exhibit 
characteristics of exo-KBs, and discuss their properties in the context 
of the evolution of our own solar system.  We also present one star that 
has a firm detection at 70~$\mu$m, but the measurement is consistent with 
the photospheric emission from the star.  In this case, we place 
unprecedentedly low upper limits on the presence of KB-like debris in 
this system.

In \S\ 2, we describe the observations and data reduction.  In
\S\ 3, we discuss the methodology for identifying the five 
exo-Kuiper Belt candidate stars from the FEPS sample, and we
present the resulting spectral energy distributions (SEDs).  
We outline our interpretation of the SEDs in terms of physical models 
in \S\ 4.  We discuss the implications of our results in \S\ 5, and 
summarize our findings in \S\ 6.

\section{OBSERVATIONS \& DATA REDUCTION}

The FEPS program is described by Meyer et al.  (2004, 2005) and a
detailed description of the data acquisition and data reduction for
the FEPS program is given in the FEPS V1.1 Explanatory Supplement
(Hines et al.  2004).  In this section, we present a condensed
description as it applies to the five targets investigated herein.
Observational details are listed for each star in 
Table~\ref{obslog}.
Imaging data were obtained with the Multi-band Imaging Photometer for
{\it Spitzer} (MIPS: Rieke et al.  2004) in the 24 \um\ and 70 \um\ bands 
using small field photometry mode with 2 -- 10 cycles of 3 and 10 
second integration times respectively; each of such dataset is a 
data collection event (DCEs).   After initial processing by 
the {\it Spitzer} Science Center (SSC) version S10.5.0 pipeline
to provide reconstructed pointing information, we used the MIPS Data
Analysis Tool (DAT) software  \citep{gor04,gor05} to process the
data.   

For the 24 \um\ data, read-2 correction, dark subtraction, droop 
correction, electronic nonlinearity correction, scan mirror dependent 
flat fields, cosmic ray rejection, and distortion correction were applied.  
We produced both individual DCE frames and a mosaic image using all DCE frames.  
Outlier rejection of each pixel was performed for a mosaic image. 
The typical number of total DCE frames per source used for photometry was
28 for 2 cycles.  

For the 70 \um\ data, dark subtraction, illumination correction, and cosmic
ray identification/rejection were applied.  The electronic nonlinearity
of the detector was corrected.  \citep{gor04,gor05} note in their
investigation that this nonlinearity is $\lesssim$ 1~\%, much
smaller than absolute calibration uncertainties of 20~\% (see the 
{\it Spitzer} Observers Manual 4.6: SOM 4.6, and the MIPS Data Handbook
2004).  A ``time-filtering'' algorithm as described by \citep{gor04,gor05} 
was used to eliminate time-dependent gain drifts.  
All 70 \um\ flux measurements were performed on the distortion corrected
final mosaic image.

We used the IDL-based software package IDP3 (Schneider \& Stobie 2002) 
to perform aperture photometry on the 24$\mu$m and 70$\mu$m data.  
IDP3 was developed by the Instrument Definition Team for the Near Infrared
Camera and Multi-object Spectrometer (NICMOS), and  has been further
optimized for the {\it Spitzer} IRAC and MIPS data.
Object fluxes were measured using standard aperture photometry techniques.
For the 24 \um\ photometry, we adopted a 14\farcs7 aperture radius and a 
background annulus from 29\farcs4 to 41\farcs7 to perform aperture 
photometry on the distortion corrected individual DCE frames.  For the 70~\um\ 
photometry we used a target aperture radius of 29\farcs7, and a background 
annulus of 39\farcs6 -- 79\farcs2 on the single mosaic image.
Aperture correction factors, listed in the SOM 4.6, were then applied.

For HD 8907, we used a smaller target aperture (19\farcs7) with the same
background annulus at 70$\mu$m to avoid contamination by a nearby source. 
We also placed a mask on the nearby source to minimize any contribution from
the wings of the PSF to the background annulus. The aperture correction for
this smaller aperture was derived relative to the standard aperture using
70um observations from other bright and isolated FEPS stars. The uncertainty
in this small-aperture correction is $\sim$ 1.2\%.
 
The 24\um\ flux uncertainties are reported in Table~2, and include both the
standard deviation of the mean flux density for individual DCEs, which is an
estimate of the internal precision of our measurement, and the absolute
calibration uncertainty.  The 70\um\ internal uncertainty is estimated by
the rms pixel-to-pixel dispersion inside the background annulus measured on
the mosaic image and scaled to the area of the target aperture.
The total uncertainty for both bands is the combination of the internal
uncertainty combined (i.e., added in quadrature) with the calibration
uncertainty.  The calibration uncertainties for 24~\um\ and 70~\um\ 
photometry are 10~\% and 20~\% respectively (see the {\it Spitzer} Observers
Manual 4.6 and MIPS Data Handbook 2004).  Note that the
the 70~\um\ internal uncertainties are much smaller than absolute
calibration uncertainty, and the internal signal-to-noise is very high for
our detected objects.  It is our knowledge of the absolute flux density in
physical units that retains the large uncertainties in Table~\ref{fluxes}.

Infrared Array Camera (IRAC: Fazio et al.\ 2004) observations at 3.6, 4.5
and 8.0$\mu$m were also obtained for each object using the
32$ \times $32 pixel subarray mode with an effective integration time of
0.01 sec per image (frame time of 0.02 s). Sixty four images of the object
were obtained at four different positions in a random dither pattern on the
array.  This gives a total of 256 images (2.56 sec total integration time)
per band.  We used the basic calibrated data (BCD) products produced by the
SSC S10.5.0 data pipeline as described in the SOM~4.6. Aperture photometry
was performed with IDP3 using a 3 pixel radius aperture centered on the
target.  The background was estimated using the 820 pixels that lie beyond
an 10 pixel radius relative to the star (i.e., all pixels beyond the target
radius). The background flux was normalized to the area of the target
aperture and subtracted from the summed target flux.  The final source flux
is the mean of the 256 measures, corrected from a 3 pixel radius to the 10
pixel radius used for the IRAC instrumental absolute flux calibration.  The
internal uncertainty was estimated as the standard deviation of the
256 independent measurements.  As for MIPS, the total uncertainty in the
physical flux density was constructed by adding the internal and absolute
uncertainties in quadrature --- absolute flux calibration uncertainties of
10~\% in all three bands.  

Low-resolution ($R = 70 - 120$) spectra were obtained with the Infrared
Spectrograph (IRS: Houck et al.\ 2004) from $7.4 - 38~$\micron, using
``high-accuracy'' blue peak-up to place the source in the
spectrograph slit.  Integration times per exposure were 6 sec over the
Short-Low wavelength range (7.4 $-$ 14.5~\um), and 14 sec over the
Long-Low wavelength range (14.0 $-$ 38.0~\um).  The spectra for wavelengths
beyond $\sim 35~\mu$m are not reliable (Houck et al.\ 2004), so have been
omitted.  The spectra were obtained at two nod positions in staring mode for
averaging and estimating the noise.  The BCDs resulting from the SSC
pipeline S10.5.0 were further processed within the SMART software package
(Higdon et al.\ 2004).  We used the {\it droopres} data products, which are
intermediate products produced before stray-light and flat-field corrections
have been applied.  Spectra were extracted assuming point source profiles
with a fiducial width of 5 $-$ 6 pixels in the center of the orders.
Residual ``sky'' emission was subtracted using adjacent pixels.  Random
errors were calculated  from the difference between the two independent
spectra, then added in quadrature with an estimated 14~\% uncertainty in
absolute flux calibration.  A more detailed discussion of the IRS data
reduction and extraction is presented in Bouwman et al.\ (2005).

\section{SPECTRAL ENERGY DISTRIBUTIONS OF FIVE EXO-KUIPER BELTS} 

As part of the ongoing analysis of the FEPS database, we have looked 
for stars that are detected at 70~\um\ but show no significant excesses above 
the stellar photospheres for $\lambda \le 33~$\micron.  This criterion 
restricts the equilibrium temperature and inner radius of a significant 
dust component to $T < 70$~K and $R_{\rm in} \geq$~10~AU for dust 
in thermal equilibrium around a star with 1 L$_\sun$.
Three other parallel investigations based on FEPS observations will 
concentrate on stars that have: 1) young optically-thick disks detected 
by IRAC, which focus on determining the evolutionary time scale of
the inner disks (Silverstone et al.\ 2005); 
2) strong emission at 24 \micron\ but weak at 70 \micron\ indicating
   material at temperatures approaching those of our solar system's asteroid
   belt and terrestrial planets (Hines et al.\ 2005); and
3) infrared excess indicating a broad range of temperatures characteristic 
   of both terrestrial zones and Kuiper Belts (Bouwman et al.\ 2005).  
The focus of this study is on stars that are dominated by excess infrared 
emission suggestive of Kuiper Belt like disks.       

Five stars were selected from the FEPS data obtained between 
May 28, 2004 (MIPS campaign 8) and Sep.\ 25, 2004 (MIPS campaign 13), 
that have $\geq$~3~$\sigma$ excess at 70~\um, but are consistent with 
photospheric emission ($<$~3~$\sigma$ excess) $\leq$~33~\um.  
Three of the objects are newly discovered debris systems (HD~6963,  
HD~145229, \& HD~206374), while two of the objects were previously 
identified as having an infrared excess (HD~8907 \& HD~122652; 
respectively, Silverstone 2000, Zuckerman \& Song 2004) from IRAS 
observations.  We note that one object, HD~8907, is one of 10 stars
with known excess sources, which was included in the FEPS targets
for the purposes of a gas emission line search.  A sixth star, 
HD~13974, was also detected at 70~$\mu$m with high signal-to-noise, 
but in this case the 70~$\mu$m flux density is consistent with emission 
from the photosphere to within 1~$\sigma$ uncertainty.  
 
\begin{figure}[t]
\figurenum{2}
\includegraphics[height=10.0cm,width=13cm]{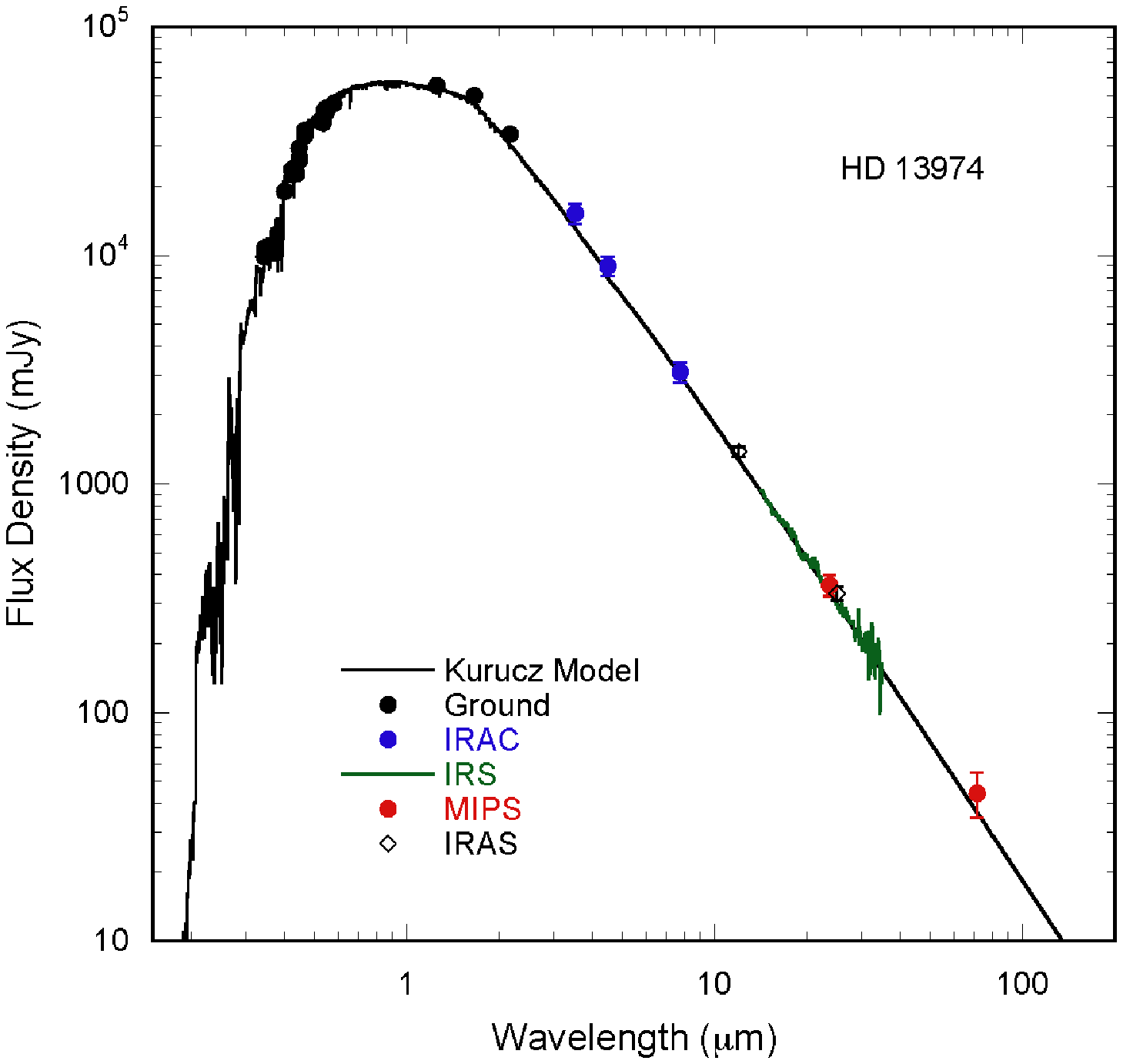}
\caption{Spectral energy distribution (SED) of HD 13974.  This system
shows photospheric emission up to 70 \um.  Symbols and lines are the
same as in Figure~\ref{4seds}.}
\label{hd13974}
\end{figure}

{\it Spitzer} photometry for the five stars is presented in
Table~\ref{fluxes}, and the SEDs are shown in Figures~\ref{4seds} and
\ref{hd13974}.  Mid-infrared spectra are also displayed for all of the
objects. We use the ``average weighted'' wavelengths of IRAC and MIPS
bands as suggested by  the \spitzer\ Observer's Manual (SOM).
Color corrections have not been applied.  The magnitude of these 
corrections are expected to be smaller than the present absolute 
calibration uncertainties  (see the IRAC and MIPS data
handbooks\footnote{http://ssc.spitzer.caltech.edu/}).

The expected photospheric emission for each star was determined by           
fitting Kurucz model atmospheres with convective overshoot to                
published optical photometry including, if available, Johnson $BV$, 
Str{\"o}mgren $vby$, Tycho $BV$, Hipparcos $H_p$, Cousins $RI$ and      
2MASS $JHK_s$ measurements.  Predicted           
magnitudes in each filter were computed by multiplying the Kurucz            
model with the combined system response of the filter, the atmospheric
transmission (for ground-based observations), and the spectral 
response function of the detector as outlined in a series of papers 
by \citet[and references therein]{co03a,co03b}.  
The Johnson-$U$ and Str{\"o}mgren-$u$ filters were excluded from the model           
fits because the observed photometry shows large deviations from the         
model values explained in part by chromospheric activity common in
young late-type stars.  The best-fit Kurucz model was computed in a 
least squares sense with the effective temperature and normalization               
constant as free parameters. 
The metallicity was fixed at [Fe/H] = 0.0 and the surface gravity
at log~g~cm~s$^{-2}$ = 4.5. In addition, each of the stars is within 
the Local Bubble (see, e.g., Welsh et al. 1998), and the visual 
extinction was fixed at $A_V = $0~mag.
The stellar parameters used in the Kurucz models, and additional 
properties of the stars including their distance and estimated ages, 
are listed in Table~\ref{stars}.  

\begin{figure}[t]
\figurenum{4}
\includegraphics[totalheight=0.5\textheight]{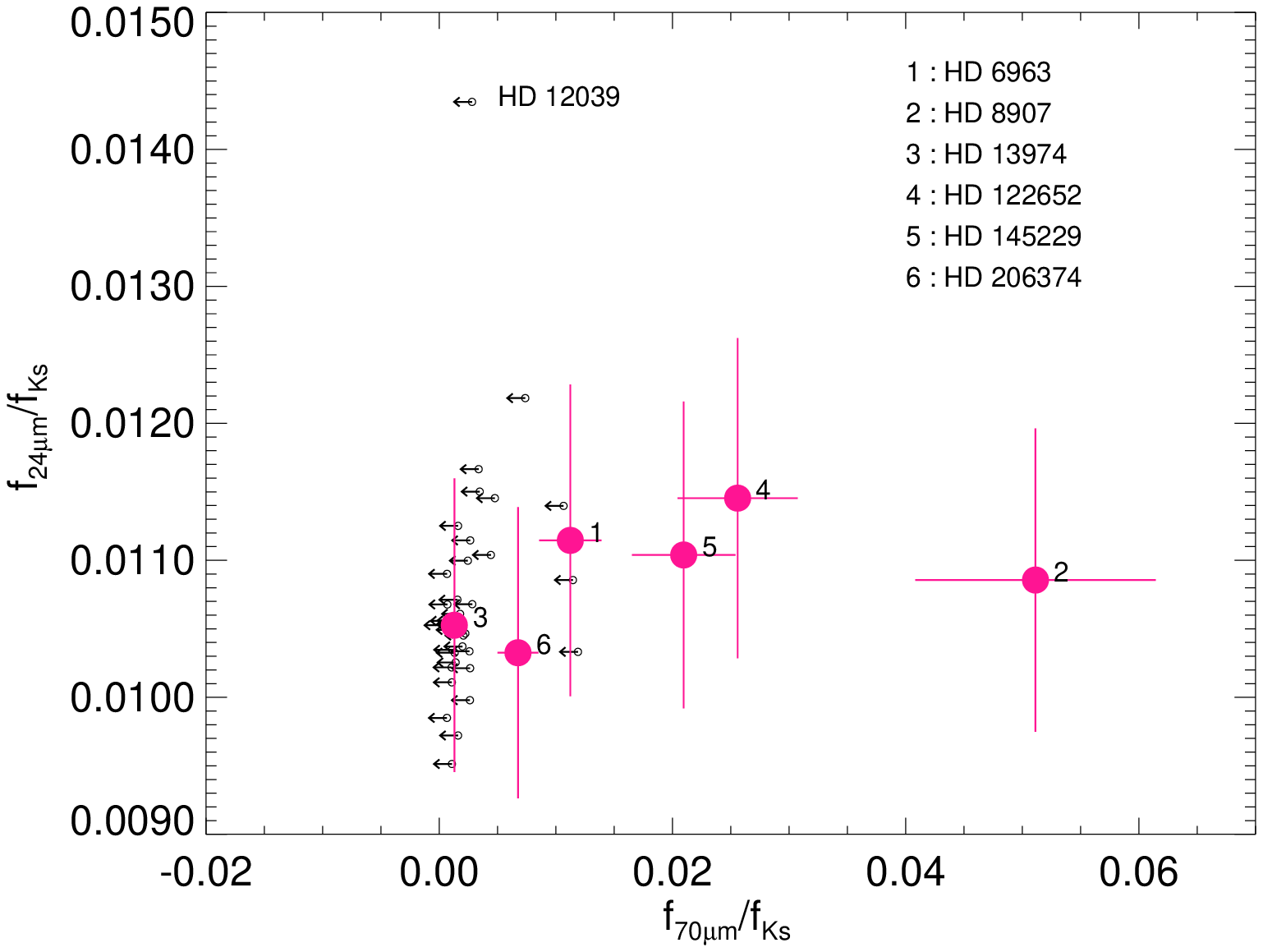}
\caption{Flux density ratio plot of 70\um/Ks vs.\  24\um/Ks. 
Small circle with left-arrow symbols are 1$\sigma$ upper limits of 
non-detections, and large filled circles marked with numbers are the 
source presented in this paper.
HD 12039 (a source with a dusty ring at around 5AU, see Hines et al.
2005) is the source with the largest 24\um/Ks value in this plot.
}
\label{ccd}
\end{figure}

The deviation from the stellar photospheres at $\lambda \sim 70~\mu$m 
is readily apparent for HD~6963, HD~8907, HD~122652, HD~145229, and 
HD~206374 (Fig.~\ref{4seds}). The 70~\micron\ emission from HD~13974 is 
consistent with the stellar photosphere (Fig.~\ref{hd13974}) within 
the uncertainty of 70~\um\ flux.  
Figure~\ref{histogram} shows the significance 
of the 70~\um\ excesses relative to the Kurucz model atmospheres for 
the five excess stars compared with the distribution for the other 
stars in the FEPS V1.1 data release.  Figure~\ref{ccd} also illustrates 
the excess in a color-color space that is sensitive to the color temperature 
of the excess (24\um/Ks vs.  70\um/Ks).  Large filled circles in 
Figure~\ref{ccd} are sources with 70~\micron\ detections, and small open
circles with arrows are 1~$\sigma$ upper limits for non-detections.
This plot has the virtue of also capturing the lack of strong 24~\um\  
excesses in these targets.  Note that the 24~\um\ excess of HD~12039 
is well separated from the other targets; this star is unique within
the FEPS program in exhibiting only warm debris dust, and is discussed
in detail in \citet{hin05}.  
The location of HD~13974 in this plot agrees well with 1~$\sigma$ upper 
limits of 70~\micron\ non-detections.  

In the next section we use the observed SEDs to constrain the
properties of the detected debris systems.  For HD 13974 the 70~\micron\
detection is consistent the photosphere, placing stringent limits on any
KB-like excess. For those objects that were not detected at 70$\mu$m, we are
still able to place upper limits on the presence of a debris system.

\section{DEBRIS DISK MODELS}
We model our debris disks assuming optically thin dust in thermal
equilibrium with the stellar radiation field.
In this case, the temperature of a dust grain with a given chemical composition
and grain size depends on the radial distance to the central star only.
The fact that the SEDs do not significantly deviate from the 
photosphere for $\lambda \le 33$~\micron\ places the maximum temperature   
for the detected dust grains at $\sim 100$~K. This suggests a minimum 
equilibrium distance from the stars $\gtrsim 10~$AU, assuming grains 
similar in size or larger than those found in our own zodiacal dust 
cloud (radius $a\sim 10 - 100$~\um, Reach et al.\ 2003). 
  
The dust grains are subject to radiation pressure and Poynting-Robertson
(P-R) drag, and the action of both mechanisms places limits on the time that
the dust remains in the system. Very small grains will be ejected from the
system by radiation pressure, while larger grains will suffer
P-R drag and spiral toward the parent star.  A detailed prescription for
computing ``blowout'' sizes and P-R drag timescales is given in Burns et
al.\ (1979) for the solar system, and are derived for a range of extrasolar
systems by Artymowicz (1988) and Backman \& Paresce (1993). Particles
smaller than $\sim 1~\mu$m have a blown out time $<$10$^2$ yrs. We consider
grains to be no longer important contributors to the SED when they travel a
factor of 4 farther from the star than their position of creation in the
main dust belt.  At these distances, the grain is 0.5 $\times$ initial
temperature, and emits at a twice the wavelength.
We consider such a grain to no longer be contributing to the main excess
SED, and in effect has left the system.  Grains blown out from points of
origins a few $\times$ 10~AU from a solar type star will leave a system by
this definition in at most only a few$\times$10$^2$ years.                   

Particles larger than $\sim$~1 \micron\ will be subject to slow P-R inward 
drift and will be destroyed on timescales of 10$^{6}-10^{7}$ yrs 
for particle sizes of $\sim$1 $-$ 10 \micron\  (timescale linearly 
proportional to particle size) starting from distances of order 
$r$ $\sim 10$~AU (timescale proportional to $r^{2}$).  
This time scale is very short compared to the age of the systems 
(see Table~\ref{stars}), which suggest that the debris in these systems 
is being replenished by a parent population of objects, a putative 
Kuiper Belt. 

From these arguments, we can build simple models of the debris               
systems. Below we develop models for five of the objects that     
have limited photometry beyond 24~\um.  For HD~8907, where we have   
more photometric constraints on the SED, we explore a more 
comprehensive model.

\subsection{Simple Blackbody Grain Models}    
Blackbody models of the detected debris disks were based on excess color
temperature ($T_{\rm c}$), estimated from the two shortest broadband
wavelengths with $\geq$ 1 $\sigma$ excess (including the synthetic band at
33 \micron\ using IRS long-low data) --- Blackbody grains by definition
absorb and emit radiation efficiently at all relevant wavelengths. For each
object, the solid angle, $\Omega$, subtended by the emitting material as
seen from the central star was calculated from $T_{\rm c}$ and excess
$F_\nu$ using the Planck formula.  This value for $\Omega$
in turn was converted to: (1) total grain emitting cross-section area,
$A_{\rm x}$, using the system distance, (2) grain luminosity using
$A_{\rm x}$ and $T_{\rm c}$ in the Stefan-Boltzmann formula for
blackbody luminosity, and (3) grain mass assuming a constant per-grain mass
density \footnote{A density of 2.5 \gcc\ was chosen to represent generic
silicate fragments of asteroid-like planetesimals, c.f., mean value of
densities determined for 10 asteroids $\sim$2.4 \gcc;
http://aa.usno.navy.mil/hilton/asteroid\_masses.htm.} of
2.5 \gcc\ for solid  ``astronomical silicate'' (Draine \& Lee 1984)
and grain radius 10~\micron\ for efficient emission at 70~\micron.
The 10~\micron\ radius grain is the smallest size that acts as a
blackbody at 70~\micron.  
The uncertainty in blackbody temperature estimate ($T_{\rm c}$) is about 10\%. 

In all cases, the flux densities between 33~$\mu$m and 70~$\mu$m are
increasing, suggesting that we are observing the Wien side of the dust 
excess SED, thus $T_{\rm c}$ is interpreted as an estimate of the maximum 
dust temperature. The corresponding minimum distance from the star of 
the emitting material \Rin\ was found from the estimated maximum 
$T_{\rm c}$ and the relevant $L_\star$ using the formula for blackbody 
grains in Backman and Paresce (1993).   
The relationship between grain temperature, position ($R_{\rm AU}$), and 
primary star luminosity (equation 3 of Backman \& Paresce 1993) for
grains larger than $a \sim 10\mu$m, that emit efficiently at $70 \mu$m,
and have negligible albedo, is
\begin{equation}
T_{\rm c} = 278 L_\star^{1 \over 4} R_{\rm AU}^{-{1 \over 2}} {\rm K}
\end{equation}
where $L_\star$ is in unit of $L_\sun$, 3.9$\times$10$^{33}$ erg s$^{-1}$.
The model fit parameters and masses are presented in Table~\ref{results},
and Figure~\ref{4seds} shows simple model SEDs.

The blackbody models also allow estimates of the mass in the radiating grains.
Because (1) these calculations made use of a maximum dust temperature and 
corresponding inner radius (\Rin), also (2) lack of photometric measurements 
or even useful upper limits beyond the apparent peak of excess emission 
(70~$\mu$m) prevents useful characterization of the dust  outer boundary 
radius, and finally (3) grains larger than the minimum size assumed here would 
radiate the same blackbody SED but would have a higher ratio of mass to surface 
area, the values of $L_{\rm IR}/L_\star$ and $M_{\rm d}$ in Table~\ref{results} 
characterize the portion of the SED observable out to 70 \micron; future 
observations at longer wavelengths could conceivably indicate more luminosity 
and dust mass.

We note that debris disk models comprised of grains slightly larger than 
the blowout size ($\sim$ 1 \micron) cannot be ruled out based on simple fits 
to the SED.  Such a model (based on a temperature vs. position function 
given in equation 5 of Backman and Paresce 1993) requires an \Rin\ 
of $\sim$ 270 AU for a solar luminosity star and a dust mass $\sim$ 3 larger 
than the lower limit given above.  We comment on the likelihood of this model 
in section 5.4.

\subsection{Detailed Modeling of HD~8907}                       

The simple black body grain model yields $T_{\rm c} = 48$~K, the inner radius 
\Rin\ = 48~AU, a dust to stellar luminosity ratio (log $L_{\rm IR}/L_\star$) 
$= -3.64$, and a dust mass log ($M_{\rm d}/M_\earth$) $= -3.08$ for 10~\um\ 
radius blackbody grains with grain density 2.5~\gcc.  

The detection of significant excess emission from HD~8907 at several
wavelengths allows us to conduct a more thorough analysis of that system.
Therefore we modeled the SED of HD~8907 using the dust disk models of 
Wolf \& Hillenbrand (2003, WH03; 2005). We used the Levenberg-Marquardt algorithm 
to solve the least-squares problem, giving a best-fit dust disk model 
(Marquardt 1963; Markwardt 2003
\footnote{http://cow.physics.wisc.edu/$\sim$craigm/idl/fitting.html}). 
A detailed description of the fitting 
algorithm will be given in Rodmann et al.\ (2005).  
Due to lack of mineralogical features in the IRS spectrum, 
we assumed the optical properties of astronomical silicates 
(Draine \& Lee 1984; Laor \& Draine 1993; Weingartner \& Draine 2001).  

We assumed the density profile to be \mbox{$n(r) \propto r^{-1}$}, 
corresponding to a disk with a constant surface density 
\mbox{$\Sigma(r) \propto r^0$}. 
The power law exponent of the grain size distribution 
\mbox{$n(a) \propto a^{-s}$} 
was set to the canonical value $s=3.5$, characteristic for a size distribution 
initially produced by a collisional cascade (Dohnanyi 1969; Tanaka et al. 1996). 
For the maximum grain size ($a_{\rm max}$) and the \Rout, we assumed arbitrary 
values of \mbox{$a_{\rm max} = 1\,$mm} and \mbox{$R_{\rm out}=100\,$AU}.
Grains larger than 1~mm contribute little to the infrared emission of the
   disks, and the outer radius is poorly constrained in the absence of
   sub-millimeter/millimeter data.
We used an upper limit at $\lambda=3.1\,$mm (Carpenter et al. 2005)
to constrain flux of the disk models at mm wavelengths.
We then fit for three parameters simultaneously in a $\chi^2$ sense:
\Rin, $a_{\rm min}$, and M$_{\rm dust}$  in grains smaller than 1~mm. 
The first two parameters influence the shape of 
the SED and the wavelength at which the disk emission begins to depart from 
the stellar photosphere; the latter only scales the dust reemission to match 
the peak of the infrared emission.

\begin{figure}[htp]
\figurenum{5}
\includegraphics[height=10cm]{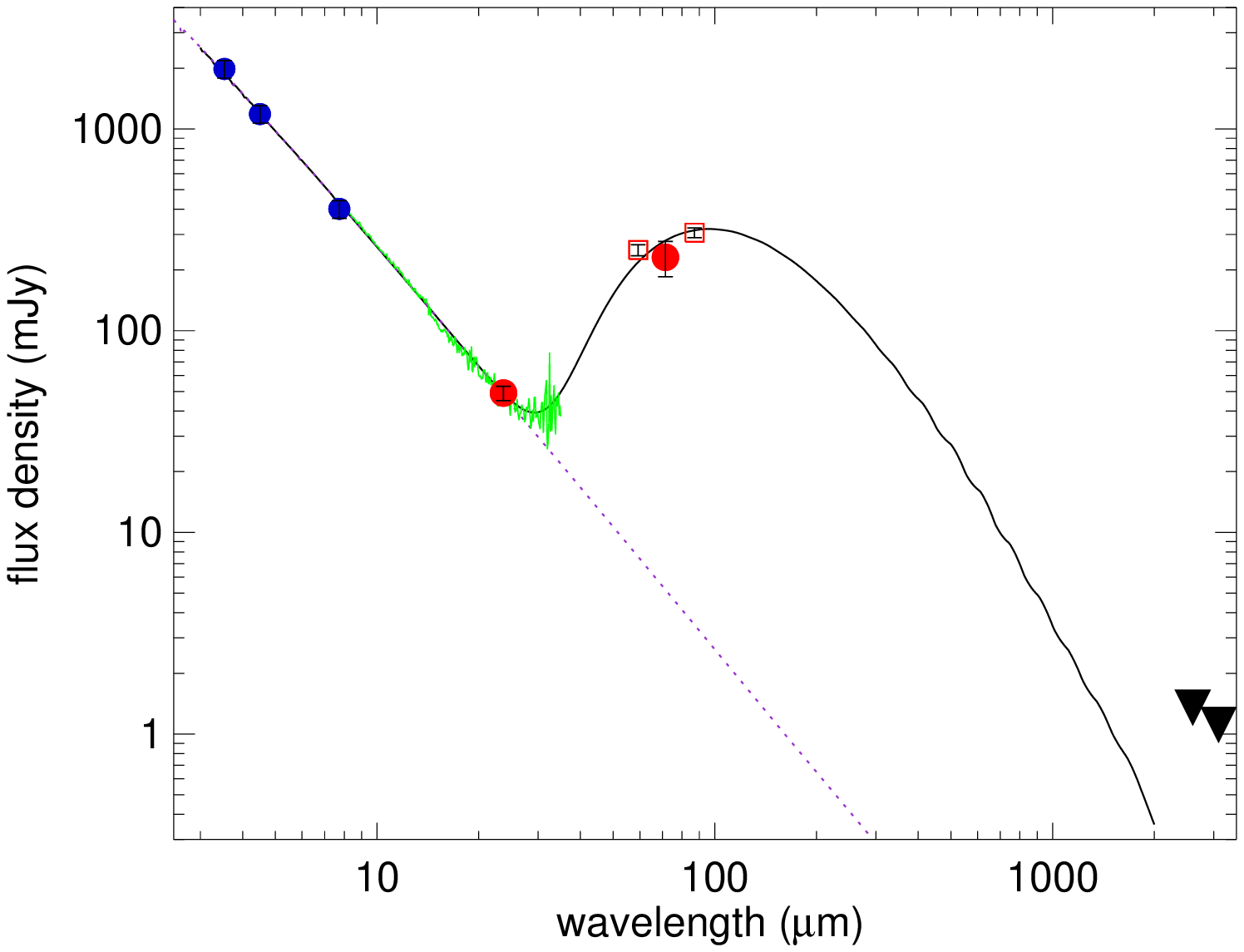}
\caption{Spectral energy distribution (SED) of HD 8907.  The model SED
is a fit from detailed modeling using WH03 models (see \S 4.2).
Blue filled circles are IRAC data, red filled circles are MIPS data,
box points are ISO fluxes, and filled upside-down-triangles are 3 $\sigma$ 
upper limits of 2.9mm and 3.1mm (Carpenter et al. 2005). 
The spectrum (green) is IRS data.}
\label{hd8907}
\end{figure}

Figure~\ref{hd8907} shows the SED of HD~8907 with the WH03 model result 
using astronomical silicate and grain radii range from 6~\micron\ $-$ 1~mm. 
The result suggests \Rin\ $\sim$ 42.5~AU, with dust mass of 1.7$\times 
10^{-2}$~$M_\earth$.           
The larger dust mass compared to the mass estimate from the 
simple blackbody model is from the inclusion of larger grains,             
which contribute the bulk of the mass even though they do not contribute    
to the bulk of the infrared emission.    
The 1 $\sigma$ uncertainty level for each parameters used in our modeling are: 
$\delta$(\Rin) $\sim$~20~AU (50\%); $\delta$($a_{\rm min}$)
$\sim$~3~\micron (50\%); and $\delta$($M_{\rm d}$) $\sim$~3.3 $\times$ 
10$^{-3}$ $M_\earth$ (20\%). 
These parameter uncertainties are estimated from the covariance
matrix (inverse of the $\chi^2$ curvature function) by 
the Levenberg$-$Marquardt algorithm.

\subsection{On the Presence of Warm Dust Mass Within $R_{in}$}                 
The selection of debris disks for this investigation is purposely           
biased against objects with debris at temperatures warmer than 
$\sim 100$~K. However, in our own solar system we observe a warm 
component associated with debris from the asteroid belt.  Such a warm 
component, albeit faint, has also been verified in other debris systems, 
e.g., around an A-star HR~4796A (e.g., Koerner et al.\ 1998), and 
recently with {\it Spitzer} imaging of Fomalhaut (Stapelfeldt 
et al.\ 2004).  We do not have the luxury of high resolution imaging 
of the five Sun-like stars presented herein, so we have to rely on 
the SEDs to place upper limits on the warm dust components.
Table~\ref{model2} presents upper limits to the amount of warm dust
located interior to the outer disk inner boundaries ($R_{\rm in}$) in each
system.  These were evaluated by assuming that dust in these locations
would be drifting in from the denser outer source zones via P-R radiation
drag and thus would extend with constant surface density from $R_{\rm in}$
inward to a radius correspond to the vaporization temperature of 
silicates at $T =$ 1500~K at $\sim$0.1~AU from the star.
For each system a surface density ($\Sigma$) upper limit was calculated
for dust between sublimation radius, $R_{\rm sub}$($T = $1500~K) and 
\Rin\ of the outer disk model such that the summed SEDs of 
inner dust model plus outer dust model plus stellar
photosphere was less than the upper limit to the observed SED, i.e.,
observations plus 1$\sigma$ uncertainty including calibration, as defined
above. This was then converted to dust mass limits assuming $a~\sim$10$\mu$m 
grains with material density of 2.5 \gcc\ as for the blackbody models of 
the cool outer disks. These masses are of order 10$^{-6}$ M$_\earth$,
the mass of a single asteroid only a few hundred km in diameter, and 
are generally 2$-$3 orders of magnitude below the lower limits for the 
masses we derive for the outer cool dust. 
Note that our \spitzer\ observations of HD~13974, which shows
photospheric emission up to 70~\um, yield an upper limit on warm dust
surface density equivalent to only about 20 times that
of the zodiacal cloud in our solar system  (i.e., 20 "zodis").

%
\section{DISCUSSION}
                                                                         
A primary goal of the FEPS project is to place our solar system in
context with other debris disk systems (Meyer et al.\ 2005).  To date,
most disk systems identified with {\it IRAS} and {\it ISO}
have been associated with either stars more luminous than the Sun
(typically A-stars), or very young systems ($t_{\rm age} \lesssim$
30~Myrs). A recent census of Sun-like stars with KB-like disks that have
been identified with {\it IRAS} and {\it ISO} has been presented by
Decin et al. (2003) and Zuckerman \& Song (2004), but these studies were
only able to identify the brightest systems and did not cover a large 
range of ages that encompass stars approaching the age of our Sun.  

The FEPS sample is a coherent effort to understand
the debris systems around Sun-like stars over a large range of ages from
3~Myrs -- 3~Gyrs. We are beginning to assemble a sufficiently large sample
of objects  over a range of ages to start to answer fundamental questions
about  the evolution of dust disks surrounding Sun-like systems with
\spitzer\ data.

\subsection{Age Determination}
The age determination of all of the stars in the full FEPS sample
is discussed in Hillenbrand et al.\ (2005).  Age bins given for the 
five stars (Table~\ref{stars}) in this study are inferred from the 
level of chromospheric and coronal activity.  The chromospheric activity 
is indicated by CaII H\&K emission.  Values of the fractional 
luminosity emitted in the H\&K lines, $R'_{HK}$, were derived by 
Soderblom (2000, private communication).  Here we adopt the calibration 
of Donahue (1993)  between log $R'_{HK}$ and age.  Coronal activity,
on the other hand, is indicated by X-rays.  The fractional X-ray luminosity, 
$R_{\rm X} = L_{\rm X}/L_{\rm bol}$, in general tracks $R'_{\rm HK}$; 
for HD~8907 in particular, relative youth is indicated by both activity 
indicators. We use coarse age bins rather than specific ages due to the 
uncertainty in the age estimates. 

\subsection{Solar System Evolutionary Model}                            
The total mass of the KB in the solar system is highly uncertain,
however some recent studies (e.g., Luu \& Jewitt 2003, Teplitz et al.\ 
1999) suggest its total mass is in the range $\sim$0.01 $-$ 1.0 M$_\earth$,
mostly in large 
planetesimal bodies.  From COBE observations at wavelengths 150~\um\ 
and 240 \um, Backman et al.\ (1995) estimated upper limits on KB dust
luminosity of $L_{\rm IR}/L_\star \lesssim$ 10$^{-6}$.  This corresponds
to a KB infrared dust mass limit of $M_{\rm d,KB} \lesssim$10$^{-5} M_\Earth$ 
assuming grains with $a =$ 10 \um\ and $\rho$ = 2.5 \gcc\ as in the present 
paper's models of exo-KBs. 
For comparison, Moro-Mart{\'i}n \& Malhotra (2003) estimate
a KB mass of about 4$\times$10$^{-6}$M$_\Earth$ in particles with
sizes between 2.4 \um\ and 160 \um\ based on flux of KB dust
detected drifting toward Jupiter's orbit (Landgraf et al.\ 2002).
We see that the systems considered in this paper contain more massive 
and luminous dust ensembles than does the KB.  But what would our 
own KB have looked like at the younger evolutionary stages represented 
by these five objects?

Backman et al.\ (2005) have devised a simple evolutionary model of the
KB based on the model of the current KB in Backman et al.\ (1995)
that can be used to estimate what our system would have looked
like at the ages, distances and stellar luminosities of our targets.
The model assumes the planetesimal population in the KB of these solar 
systems extends from \Rin\ $=$ 40~AU to 
$R_{\rm out}$ $=$ 50~AU at the ages of our targets, after migration of 
the outer planets and KB had been completed (Malhotra 1993; 
Levison et al.\ 2004), and the influence by Neptune (at 30~AU) had 
eroded the belt substantially inward of the 3:2 resonance at R $= 40$~AU.
The model also assumes that the starting mass of the KB was 10 M$_\earth$, 
the minimum necessary to yield a high enough density to build the observed 
large KB bodies accretionally in the allowed time span \citep{ste97}.  
The model evolution is not very sensitive to the starting
mass because a purely collisional system evolves asymptotically
toward a state in which the collision timescale is approximately
equal to the age of the system.  For example, a wide range of
starting masses evolve in this model to a KB with a mass of about
0.5 M$_\earth$ at age 4.5 Gyr.  This means that, in a dissipative
system such as this, evidence of the original state is mostly erased
and cannot be inferred by modeling {\it backwards}.

A parent body size of 10~km and maximum fragment size of 5~km were
parameters tuned to yield a KB dust population at an age 4.5~Gyr 
consistent with present limits on KB dust emission (Backman et al.\ 1995; 
Teplitz et al.\ 1999) and inference of the KB dust production rate from 
dust impact rates in the outer solar system (Landgraf et al.\ 2002).
The dust distribution and thermal emission were calculated for 
30 logarithmically spaced size bins between 1 \um\ and 1 mm diameter.
An equilibrium spatial distribution of dust was calculated,
as in Backman et al. (1995), balancing the dust between production in
planetesimal collisions, inward drift of dust via P-R radiation drag, 
and destruction by mutual collisions of dust yielding fragments smaller 
than the blowout size.
The dust distribution was modeled both within the planetesimal zone at
40 $-$ 50 AU and in a P-R induced ``zodiacal" dust cloud extending
inward and assumed completely truncated at \Rout\ $=$ 30 AU. 

Results of the KB evolutionary model runs are presented in Table~\ref{model3}, 
where columns 3 and 4 are the predicted and observed 70~\um\ excess 
fluxes of each source. Remarkably, although the solar system evolutionary 
model was tuned only to represent our KB, the results are within a factor 
of only 2 $-$ 3 from the targets' observed 70~\um\ excesses, except
for the HD~13974.   

The solar system KB evolution model predicts a 70~\micron\ excess of 
$\sim$ 21~mJy for the Sun observed from $d =$ 30~pc at an age of 
$t = $4.6~$\times~10^9$~yrs (on top of 5 mJy photosphere).  This predicted 
excess flux is less than any of the 5 detected excess sources, e.g.,
about 30 \% of the dust flux at 70 \micron\ observed for HD~145229 
with 0.96 L$_\sun$ observed at 33 pc.  Therefore the solar system present-day 
KB model has about 30 \% of the dust mass of the HD~145229 cold dust.  
These calculations indicate the possibility that the target systems 
represent snap shots of the history of our own solar system suggesting
that systems like our own might be common among G stars in the galactic
disk.

\subsection{HD~13974}
HD~13974 is a short period binary system (period = 10 days, Duquennoy \& 
Mayor 1988) with a G0V primary (Duquennoy \& Mayor 1988) and 
a companion with spectral type between G9V (Duquennoy \& Mayor 1988) and
K4V (Hummel et al. 1995).  
HD~13974 at 11~pc is the only star in our FEPS sample to date 
with a 70~\um\ detection that is consistent with ``bare''                  
photospheres within the calibration and Kurucz model uncertainties       
(Figure~\ref{hd13974}).  
A two-component Kurucz model was fitted to the observed SED with
derived temperatures of 6215 K and 4493 K for the primary and
secondary components respectively.  
The semi-major axis of these stars are
$a_1~sin~i$ $=$ 1.45 $\pm$ 0.03 $\times$ 10$^6$ km and
$a_2~sin~i$ $=$ 1.62 $\pm$ 0.04 $\times$ 10$^6$ km (Duquennoy \& Mayor 1988).
Such a close ($\ll$ 1 AU) binary star system should not have a dramatic effect 
on the evolution of the circumbinary disk at distances greater than roughly
twice the semi-major axis (e.g., Jensen et al. 1996; Artymowicz \& Lubow 1994 ).

Adopting the blackbody grain model we have used the 1$\sigma$ uncertainty 
in the observed flux density at 70~\um\ compared with the predicted Kurucz 
model flux to place an upper limit on the mass of dust 
log ($M_{\rm d}/M_\earth$)$ = -5.1$ at $T_{\rm c} = $ 55~K. 
The lower limit on \Rin\ is about 28~AU.  
The fractional dust luminosity log~($L_{\rm IR}/L_\star$) $< -5.2$ of this
system is interesting, as it is similar to the estimates for the 
current solar system Kuiper Belt.  
The solar system KB evolutionary model predicts 70~$\mu$m excess flux density 
of HD~13974 at its age (1 $-$ 3 Gyr) to be about 300~mJy, while the observed 
excess flux density limit is $<$ 20~mJy (Table~\ref{model3}).    
This system may not contain KB bodies, or perhaps there is not
a perturbing planet like Neptune to stir up the system and cause
collisional cascades.

\subsection{Possible Planetary Architectures in the Five IR-Excess Stars}
For the five systems in which 70~\um\ excesses have been detected, the
comparison between the disk mass estimate ($M_{\rm d}$ in
Table~\ref{results}), and the upper limit to the amount of warm dust
($M_{\rm d,warm}$ in Table~\ref{model2}) located interior to the
inner boundary of the dust (\Rin\ in Table~\ref{results}),
indicates that the dust depletion inside \Rin\ is significant.
Because grains not directly ejected by radiation pressure tend
to spiral toward the star due to P-R drag, a central depletion would
be filled in by dust on P-R timescales (Table 5) much shorter than the
ages of these systems, unless some other mechanism intervenes to
eliminate grains. 
We argue that ice sublimation and grain ``blowout" both fail to explain 
the \Rin\ locations in these systems, and that a possible explanation 
for the dust spatial distribution is a sizable planet in each system 
limiting the inward drift of the grains by ejecting them out of 
the system via gravitational scattering.
                                                                                
An upper limit of the location (\Rsub) at which grains will sublimate 
can be calculated by assuming that they are composed of water ice with 
sublimation temperature $T_{\rm sub}= $100~K.  The maximum grain 
temperatures ($T_{\rm c}$) observed in these systems presented in 
Table~\ref{results} are all well below the ice sublimation temperature. 
The presence of grains large enough to radiate efficiently at 70 $\mu$m 
implies that their temperature will scale as 
$T_{\rm c} \propto$ $R^{-0.5}$ \citep{bac93}. 
The radius of ice sublimation $R_{\rm sub}$ will be $1/4 - 1/3$  
of $R_{\rm in}$ for $T_{\rm c}$ in the range $50 - 70$ K.  
and thus sublimation is not a likely explanation for the inner  
depletion of the disk.   The uncertainties in the temperature   
estimate are about 10\% for individual sources.
  
Can these central zones that are relatively free of dust be explained as 
the result of dust grains in the outer rings being controlled
by mutual collisions rather than P-R drag?  Thus the inner edge of the dust
distribution might be explained without invoking the existence of a planet
to consume inward drifting grains.
In their study of the age dependence of Vega-like excesses, Dominik \& Decin
(2003) argued that in dense enough disks, the collision time scale is
much shorter than the P-R drag time scale.   If grain collisions actually
dominate P-R drag as the dominant dynamical process then grains could be
ground to the blowout size, and would not drift radially. 
In that case, the inner edge to the dust distribution could represent 
the inner edge of the source bodies.   
The dust collision timescales ($t_{\rm coll}$) for the  
disks can be estimated as $P_{\rm orb}$/9$\Sigma$, where $P_{\rm orb}$
is the orbital period at $R_{\rm in}$ and $\Sigma$ is the fractional surface
density of the disks that is of order the value of $L_{\rm IR}/L_\star$ 
(Backman \& Paresce 1993; Backman 2004).   
The timescales for particles to drift via P-R drag over distances equal
to 10\% of the \Rin\ values in table~\ref{results} (i.e., an estimate of 
the time for grains to move from populated to relatively unpopulated regions) 
are about log $t_{\rm P-R}~\sim$~5.5 to 6.0 years.  The grain-grain collision 
timescales for the same systems are about log $t_{\rm coll}~\sim$~5.0 to 5.5 years.
For ratios of this sort between $t_{\rm P-R}$ and $t_{\rm coll}$, 
a substantial fraction of grains would survive enough collision lifetimes 
to be able to drift from the disk into a central void.  
Note also that without knowing more about the disks (especially, their 
vertical extents) we do not know if grain-grain collisions would be at 
high enough speeds to destroy the grains. Thus we conclude that P-R drag 
is important in controlling the structures of these systems, especially 
in that a planetary barrier would be a plausible explanation for a central 
depletion.                                     

Could these cold disks arise from grains generated recently from 
planetesimals located near \Rin\ and then ejected by radiation pressure 
from the system, again without needing a planet to explain the inner 
boundary of the disk? We note that these debris belts can be self-stirred 
without needing a planet to explain the collisional cascade itself 
(e.g., also Kenyon \& Bromley 2004).
Recent \spitzer\ observations of Vega at 24, 70, and 160~\um\ show that 
the Vega debris disk has an inner boundary at $\sim$~86~AU, and extends
to distances much larger than those observed at submm and mm wavelengths
(Su et al.\ 2005).  The authors suggest that the grains in the Vega
far-IR disk are produced close to the inner disk boundary and are 
unbound, flowing away from the star.  In this scenario, the inner
gap in the disk would simply represent a limit to the location of the 
parent bodies, rather than the location of a perturbing planet.
However, Vega's luminosity to mass ratio (L/M) is 24 in solar units, 
which makes even relatively large grains (like the 18 \micron\ grains) 
to be unbound, with $\beta >$ 0.5, if the grain's porosity is $>$0.56. 
($\beta$ is the ratio of the radiation pressure force to the gravitational
force). In contrast, for the G star systems we consider here with L/M $\sim$ 1,
the 10 \micron\ silicate grains that dominate the dust emission at 70 \micron\ 
have $\beta =$ 0.02. Therefore, we argue that grains in the five disks 
discussed here are bound and as such will drift toward the star due to
P-R drag. 

One explanation for the inner empty zones is that one or more 
massive planets dynamically deplete the dust generated by an outer belt 
of planetesimals (e.g., Liou \& Zook 1999, Moro-Mart{\'i}n \& Malhotra 
2003, 2005). 
If we assume a single planet in a circular orbit about each
of the five stars, then this mechanism can account for the large
dust depletion factors inferred for the inner parts of these dust disks
if the planet is $>$ 1 M$_{\rm Jupiter}$ (Moro-Mart{\'i}n and Malhotra 2005). 
The inner edge of the dust disk would be located between 
0.8 $\times a_{\rm pl}$ and 1.25 $\times a_{\rm pl}$,  
where $a_{\rm pl}$ is the semi-major axis of the planet.  Therefore, a
possible explanation is that the dust disk surrounding each of these five
systems harbors a planet with an orbital radius of approximately
10 $-$ 20 AU, together with a belt of dust-production planetesimals exterior 
to the planet's orbit, and no asteroid belt with significant dust production
interior to the planet's orbit.

As noted in \S 4.1, the data are also consistent with a debris disk comprised of 
1 \micron\ sized grains, inconsistent with the assumptions required for blackbody 
emission discussed above, yet larger than the blowout size.  In such a model, 
the timescale for mutual collisions is more than 100 times shorter than the 
timescale for P--R drag to significantly affect the orbital radius.
This suggests that a ring such as inferred from the model could be maintained 
through mutual collisions with dust removal dominated by radiation pressure 
blowout of the smallest particles $<$ 1 $\mu$m on short timescales.  
However, this requires the formation of a sun--like star lacking an inner disk, 
with a remnant planetesimal belt capable of generating dust through mutual 
collisions only outside of 200 AU with dust only at the smallest possible stable 
particle size.  
While the presence of a low mass companion could in principle explain such a 
large gap in the circumstellar environment, no wide stellar companions are 
known in these systems, and we do not expect many brown dwarfs (due to the 
observed lack of such systems) or giant planets (given current formation 
scenarios) at these radii.
We believe that models requiring larger grains ($\sim$ 10 \micron) at smaller 
radii ($\sim$ 30 AU) with the possible presence of a planet (or planetary system) 
interior to 30 AU are favored as they are consistent with the structure and extent 
of circumstellar disks observed around pre--main sequence stars during the epochs 
when planetary systems are thought to form.  We hope to distinguish between these 
two hypotheses in the future through direct imaging of the outer edge of the disks 
in scattered light and placing strong constraints on the presence of very low mass 
companions at large separations (e.g. with NICMOS coronagraphy on HST and 
ground--based AO observations). 
                                                       
\section{SUMMARY}  

We have presented {\it Spitzer} observations of five Sun-like stars 
that possess Kuiper Belt-like debris systems (HD~6963, HD~8907, HD~122652, 
HD~145229, and HD~206374).  Of these two were previously suggested to have 
debris disks (HD~8907 and HD~122652), and three are newly discovered. 
The fractional luminosity ($L_{\rm IR}/L_\star$) 
and ages of these newly identified disks illustrates the potential of 
the FEPS program to measure fainter debris systems, which will allow 
a more detailed census of their nature and evolution.  

We summarize our results:

1. The five excess sources have SEDs that are consistent with photospheric
models out to 33\um, but show clear excesses at 
70~\um, which was the selection criterion.  We find that these stars are 
all "old"  (four sources are in our 1 $-$ 3 Gyr age bin, while one, 
HD 145229, is in the 0.3 $-$ 1 Gyr age bin).  

\begin{figure}[t]
\figurenum{3}
\includegraphics[totalheight=0.5\textheight]{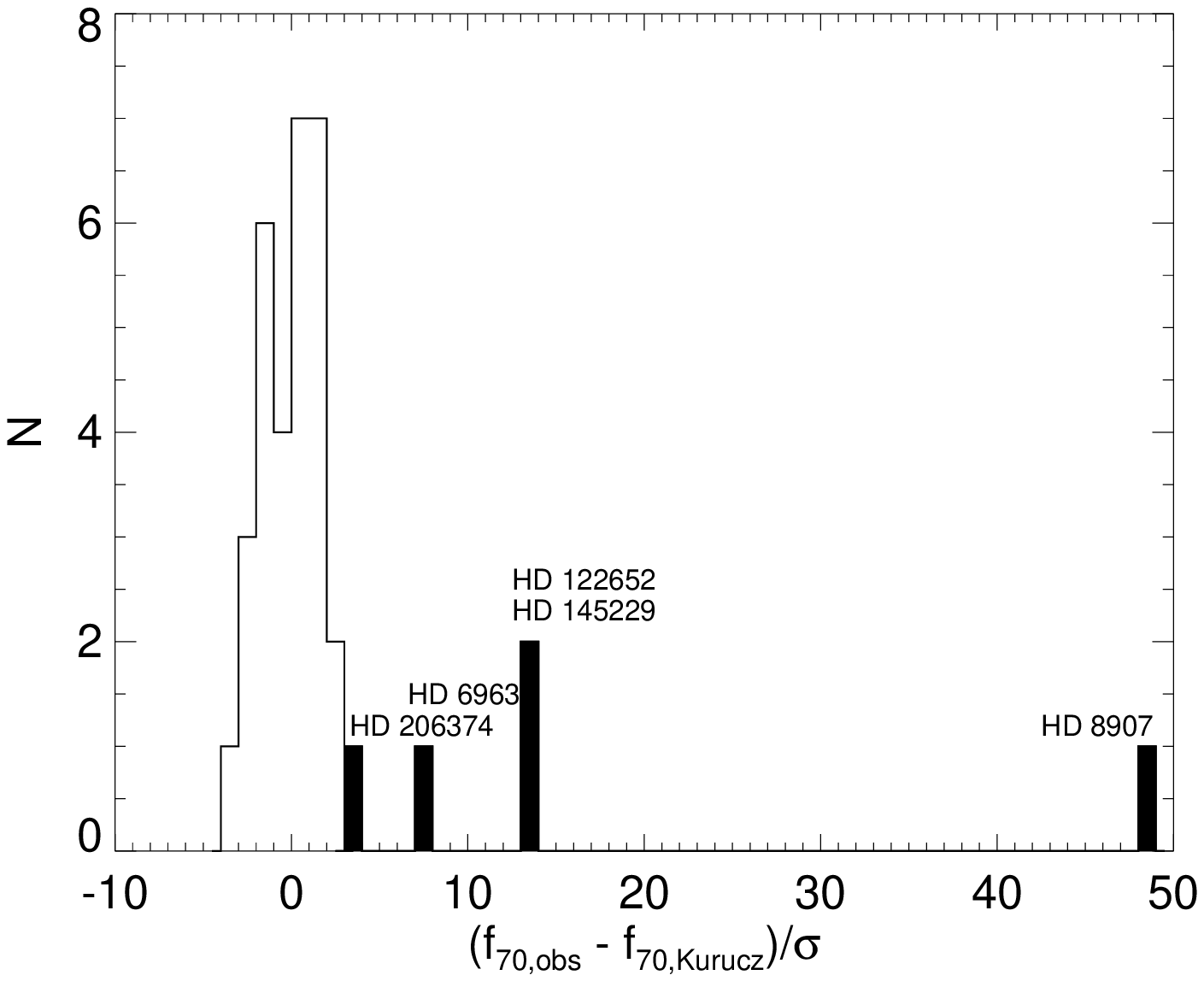}
\caption{Histogram of (observed -- photospheric 70 \um\ fluxes) divided
by $\sigma$(random uncertainty).  Four excess sources (filled histogram) 
are easily noticed in this histogram.  Note that HD 8907 shows very strong 
excess flux at 70 \um, which was detected by IRAS and ISO. HD~6963, HD~122652,
HD~145229, and HD 206374 discovered by FEPS program are nearer to the rest 
of the stars without IR excess.}
\label{histogram}
\end{figure}

2. As seen in Figure~\ref{histogram}, the improved sensitivity
of \spitzer\ allows us to detect debris disk systems that are much
fainter than those detected by IRAS and ISO.
The overall impression is that KB-like systems detectable by Spitzer 
and considered in this paper are less massive and more distant
than systems detected with IRAS and ISO.

3. Another star, HD~13974, has a MIPS 70 \um\ flux consistent with 
photospheric emission within 1 $\sigma$ total uncertainty.  
The upper limit of log($L_{\rm IR}/L_\star$) is $< -5.2$, 
similar to that of inferred for the solar systems' KB. 

4. Simple blackbody grain modeling of our 5 excess SEDs yielded 
log($L_{\rm IR}/L_\star$) $\lesssim -4.5 - -3.5$, color temperatures 
between 55 $-$ 58~K, and inner radii of outer disks between 18 and 46 AU.

5. A solar system KB evolution model predicts \spitzer\ 70 \um\ fluxes 
(Table~\ref{model3}) from hypothetical planetesimal assemblages 
around our target stars that are within factors of 
$\sim$2 $-$ 3 of the observed fluxes.  
We infer that these systems have 
outer remnant planetesimal belts that are consistent in scale and 
starting masses to our Kuiper Belt.

6. 
The absence of a disk around the $\sim 1$~Gyr 
old star HD~13974 suggests that either this object does not contain 
the parent bodies that produce infrared-emitting debris, or perhaps 
the debris has been cleared out already.

7. We placed upper limits on warm dust masses interior to \Rin\
for each of these systems, and showed that the depletion of the 
disk $<$\Rin\ is significant.  We commented on several possible causes 
for \Rin.  
We speculate that the \Rin\ of exo-KBs presented in this study could
be explained by the existence of one or more Jupiter mass 
planets at 10 $-$ 20 AU from each star.

\acknowledgments
We thank to the rest of the FEPS team members and {\it Spitzer} Science 
Center help desk. 
We have used the SIMBAD database.  This work is based [in part] on
observations made with the Spitzer Space Telescope, which is operated
by the Jet Propulsion Laboratory, California Institute of Technology
under NASA contract 1407.  FEPS is pleased to acknowledge support
through NASA contracts 1224768, 1224634, and 1224566 administered
through JPL.  S.W. was supported by the German Research Foundation 
(DFG) through the Emmy Noether grant WO 857/2-1.  EEM is supported
by a Clay Fellowship from the Smithsonian Astrophysical Observatory.
MPIA team is supported through the European Planet Network.

\clearpage
%
%
%

\begin{deluxetable}{lcccccc}
\tabletypesize{\scriptsize}
\tablewidth{0pt}
\tablecolumns{7}
\setlength{\tabcolsep}{0.06in}
\tableheadfrac{0.1}
\tablecaption{Dates and Modes of Observations\label{obslog}}
\tablehead{
\colhead{Source}        & \multicolumn{2}{c}{IRAC} & 
\multicolumn{2}{c}{IRS} & \multicolumn{2}{c}{MIPS} \\ 
\colhead{} &              \colhead{Obs. Date} & \colhead{AOR key} & 
\colhead{Obs. Date} & \colhead{AOR key} &  \colhead{Obs. Date} & \colhead{AOR key} 
}
\startdata
HD 6963    & 8/06/04 & 5387776 & 8/07/04  & 5395968 & 8/18/04 & 5396224 \\
HD 8907    & 7/25/04 & 5361920 & 1/02/05  & 5362176 & 8/23/04 & 5362432 \\
HD 13974   & 8/14/04 & 5410816 & 8/07/04  & 5411072 & 8/27/04 & 5411328 \\
HD 122652  & 6/29/04 & 5427712 & 6/06/04  & 5427968 & 6/22/04 & 5428224 \\
HD 145229  & 7/19/04 & 5387264 & 7/17/04  & 5387520 & 8/06/04 & 5387776 \\
HD 206374  & 7/04/04 & 5414656 & 6/06/04  & 5414912 & 7/11/04 & 5415168 \\
\enddata
\end{deluxetable}

\begin{deluxetable}{lcccccc}
\tabletypesize{\scriptsize}
\tablewidth{0pt}
\tablecolumns{7}
\setlength{\tabcolsep}{0.06in}
\tablecaption{Spitzer Photometry in Milli-Jy \label{fluxes} \tablenotemark{a}}
\tablehead{
\colhead{Source} & \colhead{3.6\um} & \colhead{4.5\um} & 
\colhead{8.0\um} & \colhead{24 \um}  & \colhead{33\um \tablenotemark{b}} & 
\colhead{70 \um} 
}
\startdata
HD 6963   &  1263  $\pm$126  &  729 $\pm$73  &  255 $\pm$26  &  31 $\pm$3  &  19 $\pm$  3 &  32 $\pm$7  \\ 
HD 8907   &  1985  $\pm$199  & 1186 $\pm$119 &  402 $\pm$40  &  49 $\pm$5  &  41 $\pm$  6 & 231 $\pm$46 \\ 
HD 13974  &  15274 $\pm$1527 & 8970 $\pm$897 & 3079 $\pm$308 & 360 $\pm$36 & 186 $\pm$ 28 &  44 $\pm$10 \\ 
HD 122652 &  1307  $\pm$131  &  771 $\pm$77  &  265 $\pm$27  &  34 $\pm$3  &  21 $\pm$  3 &  67 $\pm$14 \\ 
HD 145229 &  1167  $\pm$117  &  695 $\pm$70  &  237 $\pm$24  &  29 $\pm$3  &  22 $\pm$  3 &  56 $\pm$12 \\
HD 206374 &  1445  $\pm$145  &  842 $\pm$84  &  292 $\pm$29  &  34 $\pm$3  &  18 $\pm$  3 &  22 $\pm$6  \\
\enddata
\tablenotetext{a} {Uncertainties include both internal and absolute calibration 
   uncertainties. See \S 2 for more discussion.} 
\tablenotetext{b} {33\um\ synthetic photometry from IRS low resolution spectra} 
\end{deluxetable}
\begin{deluxetable}{lllcrccc}
\tabletypesize{\scriptsize}
\tablewidth{0pt}
\tablecolumns{8}
\setlength{\tabcolsep}{0.06in}
\tablecaption{Adopted Stellar Properties \label{stars}} 
\tablehead{
\colhead{Source} & \colhead{RA}      & \colhead{DEC}     & 
\colhead{Dist.}  & \colhead{Age} & \colhead{SpT} &
\colhead{T$_{eff}$} & \colhead{L}     \\
\colhead{}       & \colhead{(J2000)} & \colhead{(J2000)} & 
\colhead{(pc)}   & \colhead{(Gyr)}  & \colhead{} &
\colhead{(K)} & \colhead{(L$_\odot$)}  
}
\startdata
%
HD 6963   & 01:10:41.91 & $+$42:55:54.50  & 27  &   1--3 &  K0   & 5517 & 0.55  \\ 
HD 8907   & 01:28:34.35 & $+$42:16:03.70  & 34  &   1--3 &  F8   & 6250 & 2.05  \\ 
HD 13974  & 02:17:03.23 & $+$34:13:27.32  & 11  &   1--3 &  G0/(G9$-$K4)\tablenotemark{a}
& 6215/4493\tablenotemark{b} & 1.18\tablenotemark{c}  \\  
HD 122652 & 14:02:31.63 & $+$31:39:39.09  & 37  &   1--3 &  F8   & 6157 & 1.55   \\
HD 145229 & 16:09:26.63 & $+$11:34:28.25  & 33  & 0.3--1 &  G0   & 5893 & 0.96   \\ 
HD 206374 & 21:41:06.19 & $+$26:45:02.25  & 27  &   1--3 &  G6.5 & 5580 & 0.74\tablenotemark{c}  \\
\enddata
\tablenotetext{a} {References: Duquennoy \& Mayor (1988), Hummel et al.\ (1995)}
\tablenotetext{b} {see \S 5.3 for more discussion.}
\tablenotetext{b} {Stellar luminosity was estimated from $V$mag, distance, and bolometric 
correction for the star's temperature and spectral class.}  
\end{deluxetable}

\clearpage
\begin{deluxetable}{lcrrrrrcl}
\tabletypesize{\scriptsize}
\rotate
\tablewidth{0pt}
\tablecolumns{9}
\setlength{\tabcolsep}{0.06in}
\tablecaption{Derived circumstellar disk properties and comparison 
with other systems. \label{results}}
\tablehead{
\colhead{Source} & 
\colhead{$a$\tablenotemark{a}}  & 
\colhead{$R_{\rm in}$\tablenotemark{b}} & 
\colhead{log ($M_{\rm d}$/$M_\earth$)\tablenotemark{c}} & 
\colhead{log ($L_{\rm IR}$/$L_\star$)\tablenotemark{d}} & 
\colhead{$T_{\rm c}$\tablenotemark{e}} &  
\colhead{distance}   & \colhead{SpT}   & \colhead{References} \\
\colhead{}           & \colhead{(\um)} & 
\colhead{(AU)} &
\colhead{}& \colhead{}      & \colhead{(K)} & 
\colhead{(pc)}   & \colhead{}  & \colhead{} 
}
\startdata
HD 6963     &   10              &  18  & $-4.9$ & $-4.17$  & 57    & 33 & G0   & This study \\
HD 8907   & 10 &  48\tablenotemark{f}  & $-3.1$ & $-3.64$  & 48    & 34 & F8   & This study, Silverstone (2000) \\
HD 13974\tablenotemark{g}  & 10 &$>$28 &$<-5.1$ &$<-5.16$  & 55 & 22 & G0V  & This study, photosphere \\
HD 122652   &   10              &  31  & $-3.9$ & $-4.10$  & 56    & 37 & F8   & This study \\
HD 145229   &   10              &  24  & $-4.5$ & $-3.97$  & 56    & 33 & G0   & This study \\
HD 206374   &   10  &$>$20 & $-4.7$ &$<-4.50$\tablenotemark{h} & 57 & 27 & G6.5 & This study \\ 
\enddata
\tablenotetext{a} {grain radius used for blackbody models}
\tablenotetext{b} {Inner disk radius}
\tablenotetext{c} {Dust mass in debris disk}
\tablenotetext{d} {Fractional luminosity ratio}
\tablenotetext{e} {$T_{\rm c}$ is color temperature estimated from 33\um/70 \um\ flux ratio. 
                   Uncertainty in $T_{\rm c}$ is about 10\%.}
\tablenotetext{f} {\Rin\ $=$ 42.5~AU from more sophisticated debris disk modeling (see \S~4.2).}
\tablenotetext{g} {The 70\um\ flux of HD~13974 is consistent with photospheric emission within 
                   a total (absolute calibration and random) uncertainty.}
\tablenotetext{h} {For HD 206374 $T_{\rm c}$ is an upper limit. 
   $T_{\rm c}$ upper limit and known excess flux at 70~\um\
   correspond to lower limits on $M_{\rm d}$, because  
   more dust would be needed to equal the 70 \um\ flux if the actual
   $T_{\rm c}$ is lower than 57~K, but to an upper limit on $L_{\rm IR}$/$L_\star$
   because that scales as $T^4$.} 
%
%
\end{deluxetable}

\newpage
\begin{deluxetable}{lccrccrccc}
\tabletypesize{\scriptsize}
\tablewidth{0pt}
\tablecolumns{10}
\setlength{\tabcolsep}{0.06in}
\tablecaption{Blowout sizes, P-R drag time scales and warm dust 
(M$_{\rm d}$) mass upper limits interior to \Rin.  \label{model2}} 
\tablehead{
\colhead{Source}                              & 
\colhead{$a_{\rm blow,Si}$}                   & \colhead{$a_{\rm blow,icy}$}        &
\colhead{$R_{\rm in}$}                        &                 
\colhead{log $t_{\rm P-R,Si}$}                 & \colhead{log $t_{\rm P-R,icy}$}      &
\colhead{constraint}                          & 
\colhead{log ($M_{\rm d,warm}$)\tablenotemark{a}}   \\
\colhead{}       & \colhead{(\um)}    & \colhead{(\um)}         & 
\colhead{(AU)}   & \colhead{(yr)}     & \colhead{(yr)}          & \colhead{}       &
\colhead{(M$_\earth$)} 
}
\startdata
%
HD 6963        & 0.35 & 1.1 & 18 & 7.0 & 6.5 &  IRS 33 & $-6.1$  \\ 
HD 8907        & 0.85 & 2.7 & 48 & 6.5 & 6.0 &  IRS 33 & $-5.8$  \\    
HD 13974       & 0.58 & 1.9&$>$28& 7.0 & 6.5 & MIPS 70 & $-5.6$  \\  
HD 122652      & 0.68 & 2.2 & 31 & 6.8 & 6.3 &  IRS 33 & $-5.5$  \\
HD 145229      & 0.49 & 1.6 & 24 & 7.0 & 6.5 &  IRS 33 & $-5.7$  \\ 
HD 206374      & 0.43 & 1.4&$>$20& 6.9 & 6.4 &  IRS 33 & $-5.9$  \\ 
%
\enddata
\tablenotetext{a} {Total dust mass assuming grain radius of 10 \um\ at {R $<$ R$_{\rm in}$},
                   scaled by material density relative to 2.5 \gcc.}
\end{deluxetable}
%
\begin{deluxetable}{lccc}
\tabletypesize{\scriptsize}
\tablewidth{0pt}
\tablecolumns{4}
\setlength{\tabcolsep}{0.06in}
\tablecaption{Predicted 70 \um\ fluxes from solar system evolutionary models
and observed 70 \um\ fluxes. \label{model3}} 
\tablehead{
\colhead{Source}    & \colhead{Age}           &
\colhead{$f_{70}$(Toy S.S.)\tablenotemark{a}} & 
\colhead{$f_{70}$(Obs.)\tablenotemark{a}} \\
\colhead{}       & \colhead{(Gyr)}  & 
\colhead{(mJy)}  & \colhead{(mJy)} 
}
\startdata
%
HD 6963        &  1.0--3.0  &  70      &  30  \\ 
HD 8907        &  1.0--3.0  & 170      & 225  \\    
HD 13974       &  1.0--3.0  & 300      &$<$20 \\  
HD 122652      &  1.0--3.0  &  20      &  65  \\
HD 145229      &  0.3--1.0  &  70      &  55  \\ 
HD 206374      &  1.0--3.0  &  58      &  22  \\ 
%
\enddata
\tablenotetext{a} {excess fluxes}
\end{deluxetable}

%
\clearpage
\begin{figure}[htp]
\figurenum{1}
\includegraphics[height=10.0cm,width=13cm,angle=0]{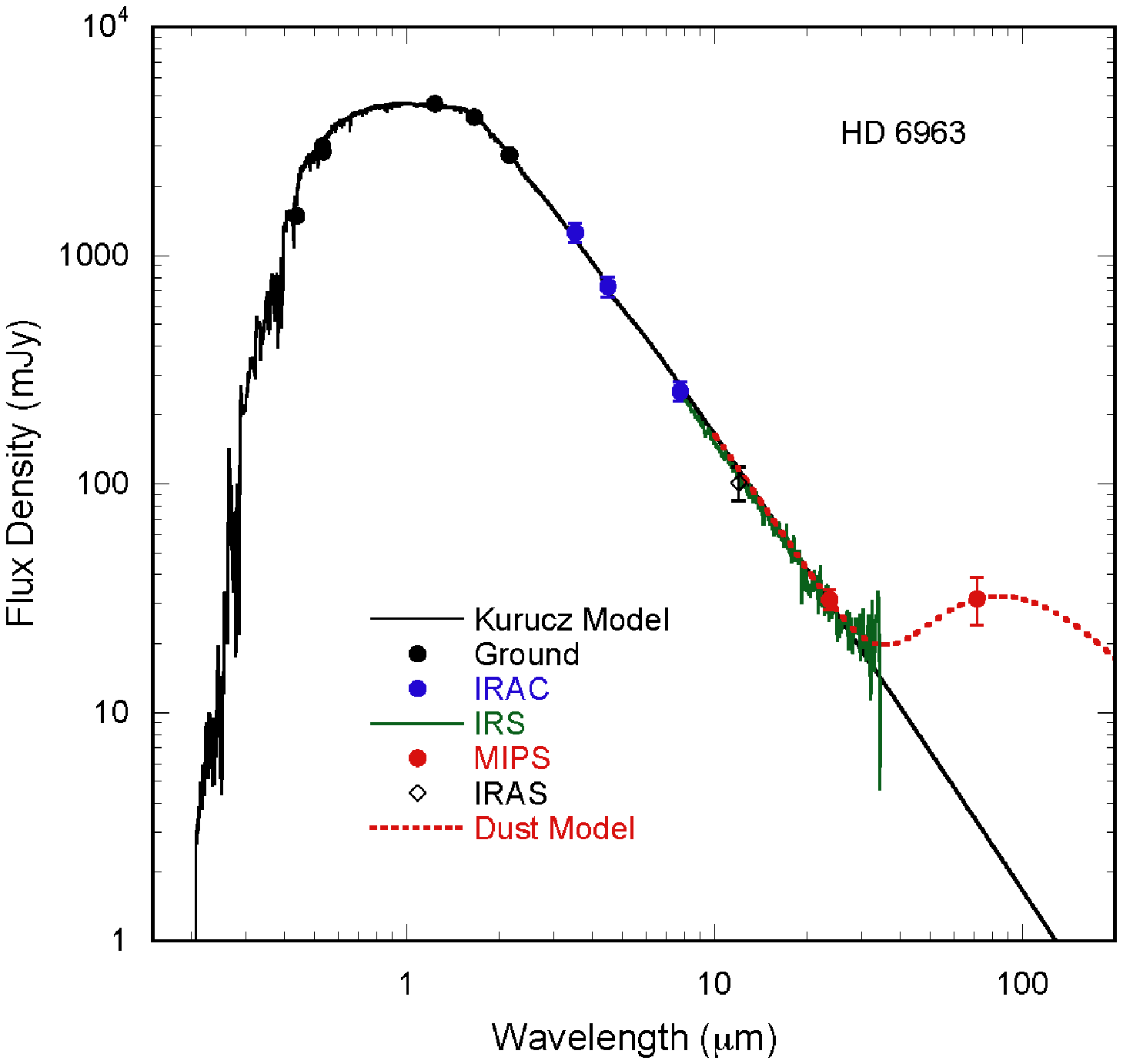}
\includegraphics[height=10.0cm,width=13cm,angle=0]{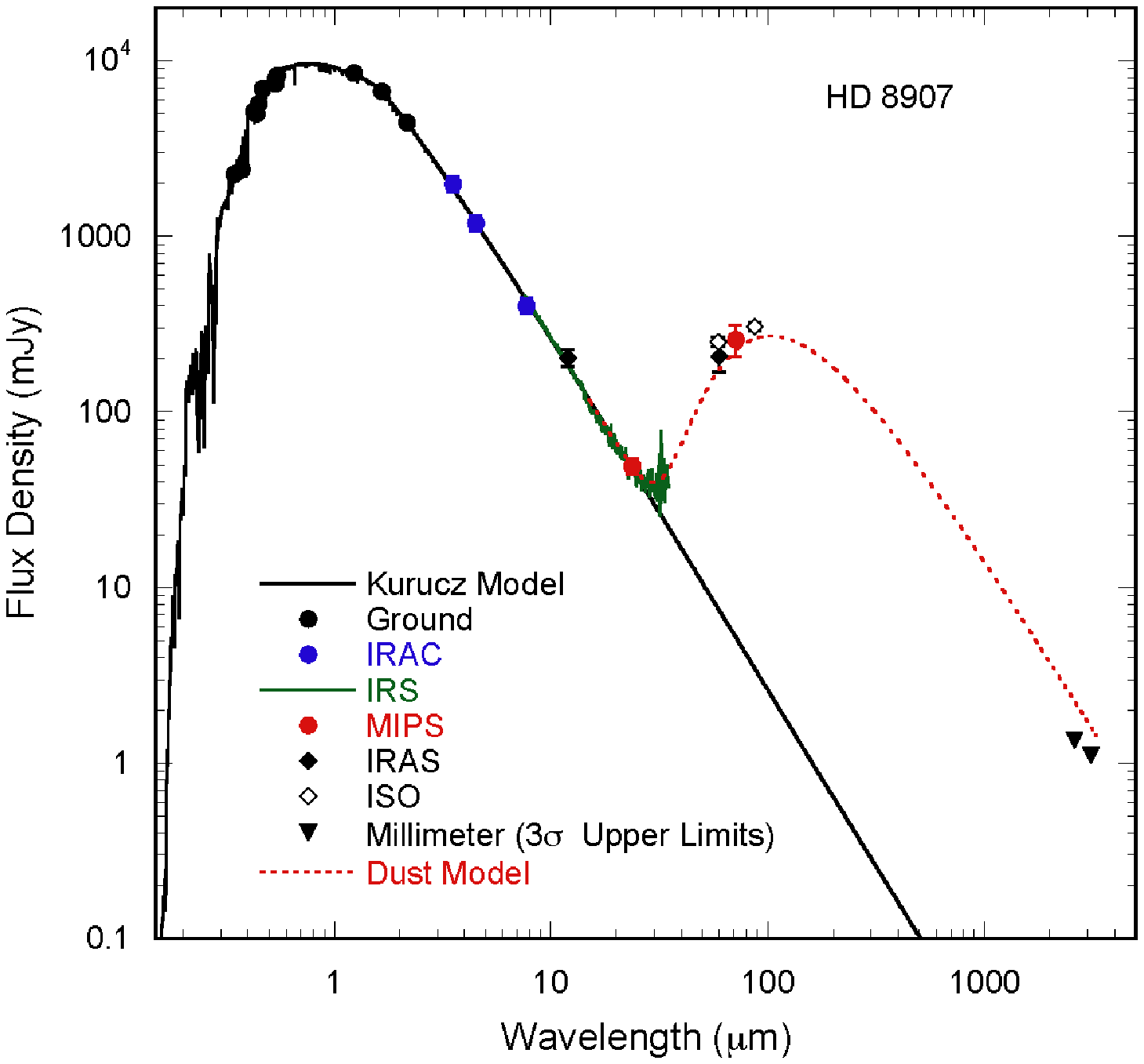}
\caption{
}
\label{4seds}
\end{figure}
\clearpage
\begin{figure}[htp]
\figurenum{1}
\includegraphics[height=10.0cm,width=13cm,angle=0]{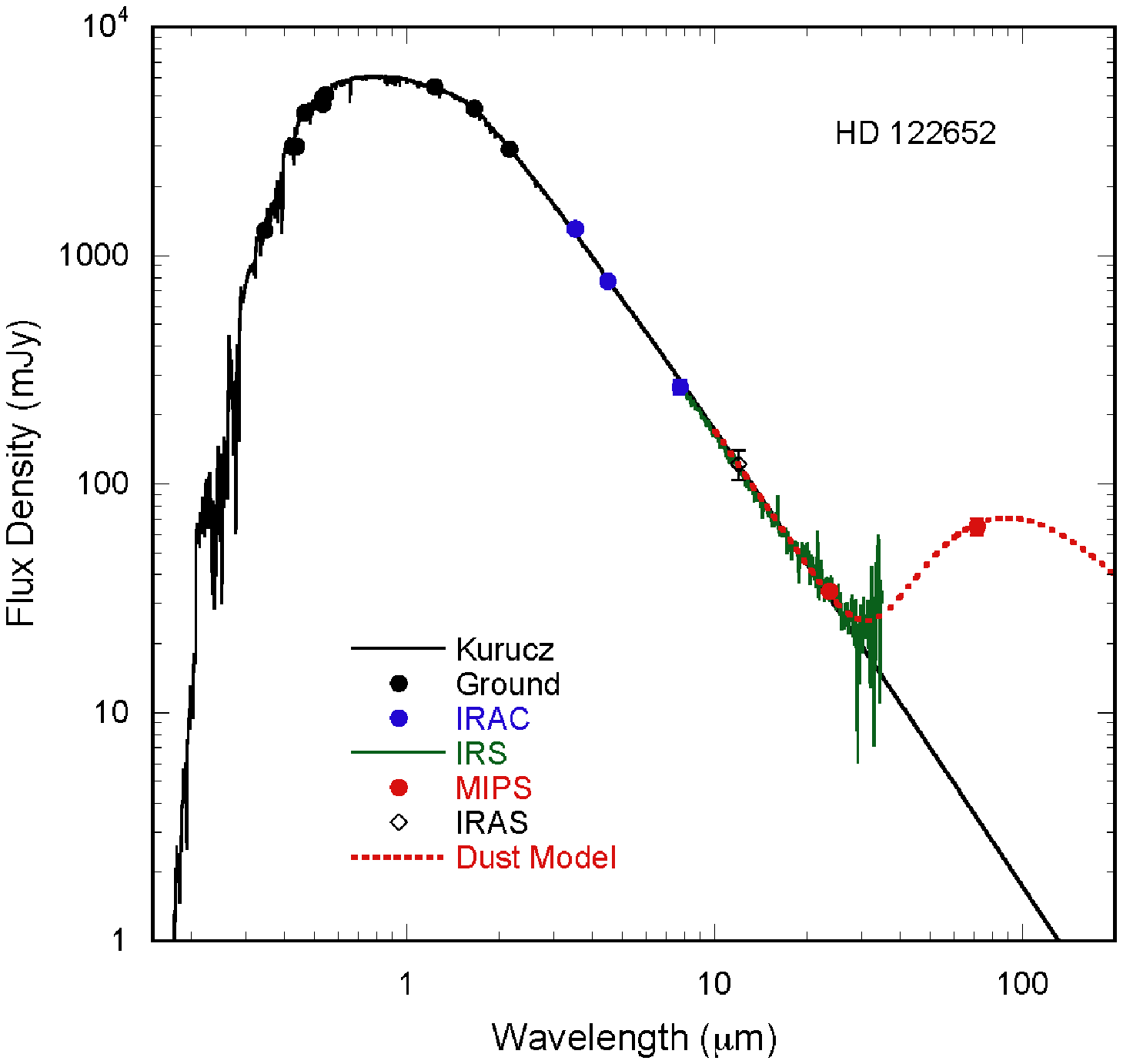}
\includegraphics[height=10.0cm,width=13cm,angle=0]{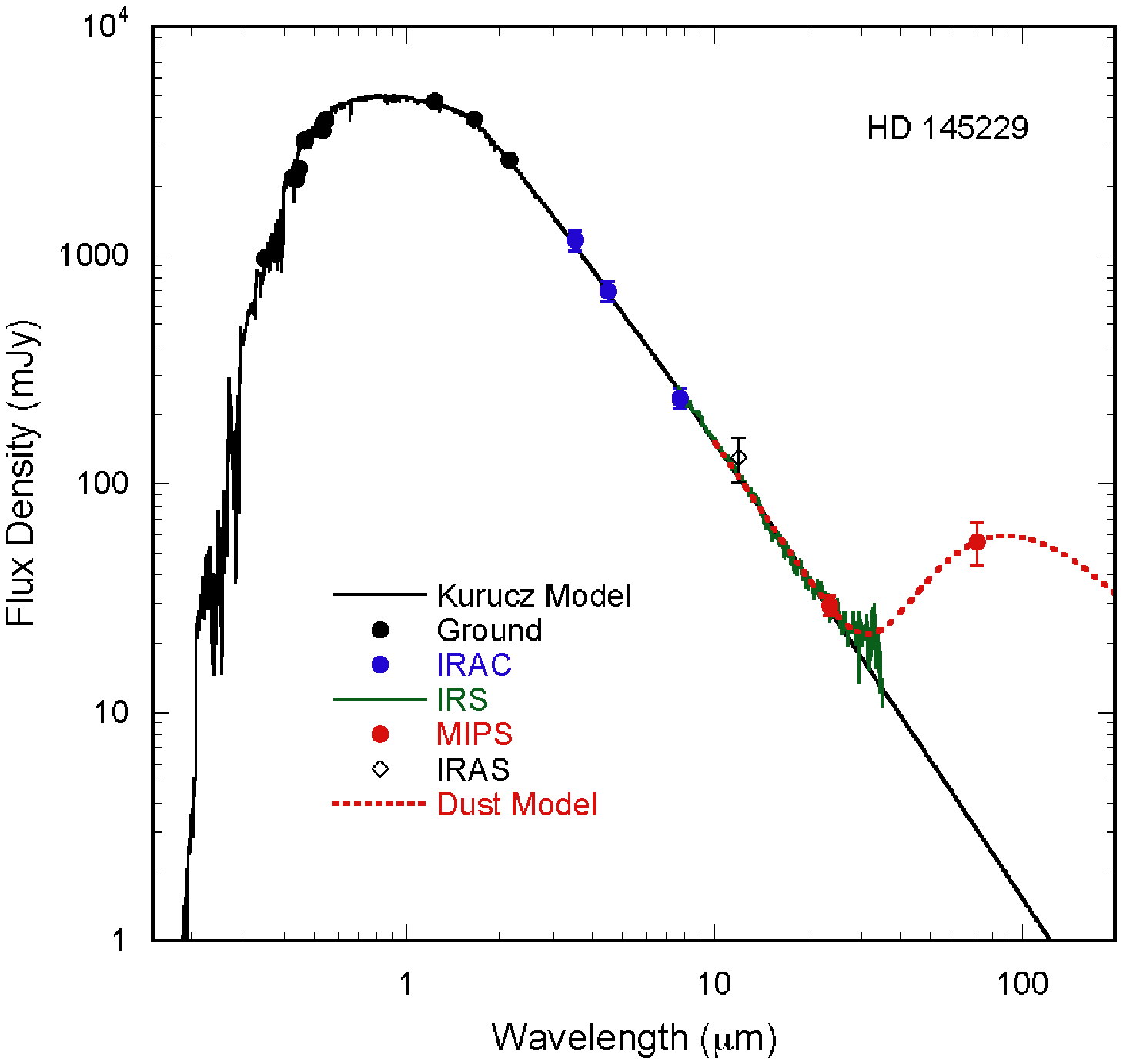}
\caption{}
\label{4seds}
\end{figure}
\clearpage
\begin{figure}[htp]
\figurenum{1}
\includegraphics[height=10.0cm,width=13cm,angle=0]{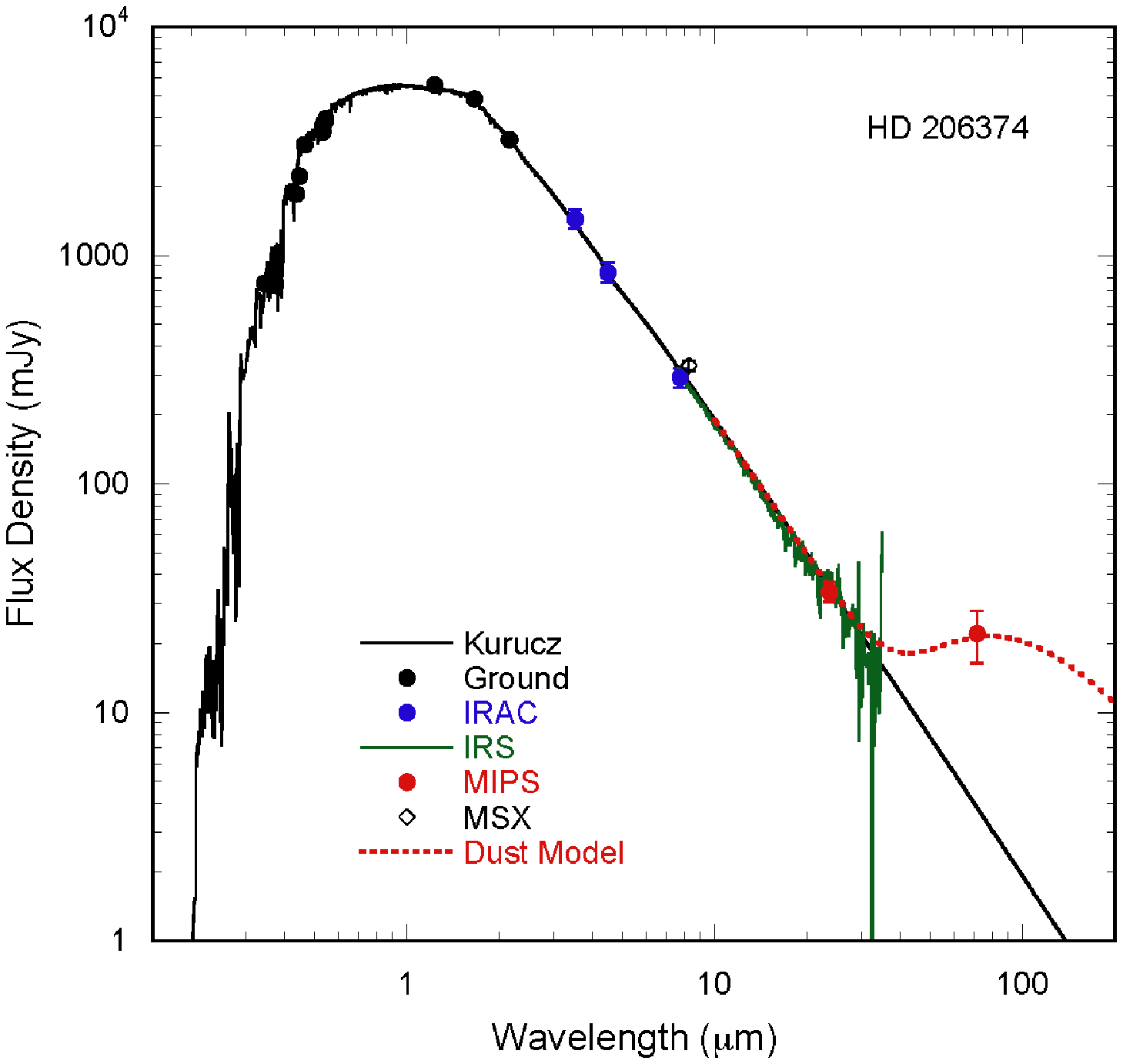}
\caption{Spectral energy distributions (SEDs) of five systems with
outer debris disks. HD 6963, HD 8907, HD 122652, HD 145229, and
HD 206374.  Blue filled-circles are IRAC data points, green
lines are IRS low resolution spectra, and red filled-circles are MIPS
data points.  Black line is Kurucz model.  Red dotted lines are
examples of possible good fit models (Backman et al.\ 2005, in prep.).
}
\label{4seds}
\end{figure}
\clearpage
\clearpage

\clearpage
\clearpage
%

\end{document}